\documentclass[prl, twocolumn, preprintnumbers, floatfix, showpacs]{revtex4}
\usepackage{graphicx}

\begin{document}
\preprint{LA-UR-13-21407 }

\title{Experimental Realization of Josephson Junctions for an Atom SQUID}
\author{C. Ryu, P. W. Blackburn, A. A. Blinova, and M. G. Boshier}
\affiliation{P-21, Physics Division, Los Alamos National Laboratory, Los Alamos, New Mexico 87545}


\begin{abstract}
We report the creation of a pair of Josephson junctions on a toroidal dilute gas Bose-Einstein condensate (BEC), a configuration that is the cold atom analog of the well-known dc superconducting quantum interference device (SQUID).  We observe Josephson effects, measure the critical current of the junctions, and find dynamic behavior that is in good agreement with the simple Josephson equations for a tunnel junction with the ideal sinusoidal current-phase relation expected for the parameters of the experiment.  The junctions and toroidal trap are created with the painted potential, a time-averaged optical dipole potential technique which will allow scaling to more complex BEC circuit geometries than the single atom-SQUID case reported here.  Since rotation plays the same role in the atom SQUID as magnetic field does in the dc SQUID magnetometer, the device has potential as a compact rotation sensor. 
\end{abstract}

\maketitle

The dc superconducting quantum interference device (SQUID) \cite{Clarke2004} was created to study and utilize the quantum interference of currents flowing in parallel through Josephson junctions (JJs) connected in a superconducting loop.  Today SQUIDs are the basis of some of the most sensitive magnetometers \cite{Clarke2004}.  An ``Atom SQUID'' is an analogous quantum interference device that uses a superfluid dilute gas Bose-Einstein condensate (BEC) \cite{Cornell1, Ketterle1} flowing through potential barriers in a trap.  The superfluid counterpart of the SQUID response to magnetic fields is an analogous response to rotation, already demonstrated in the laboratory with superfluid helium SQUIDs \cite{Schwab1997, Avenel1997, Sato2012}.  The atom SQUID is particularly attractive for rotation sensing and for addressing basic questions in quantum physics because of the BEC advantages of easy detection of atom number and phase, along with the existence of accurate microscopic theories.  

An atom SQUID has two essential ingredients: a multiply connected geometry trap (e.g. a toroid) with quantization of superfluid circulation, and potential barriers (“junctions”) exhibiting Josephson effects (i.e. tunnel junctions or weak links).  Atom SQUID counterparts of both types of SQUID are possible: a ``dc Atom SQUID'' with two junctions, and an ``rf atom SQUID'' with a single junction.   BECs have been created in toroidal traps using a magnetic trap with a repulsive optical trap at the center \cite{Ryu2007}, a Laguerre-Gaussian mode of a laser beam \cite{Ramanathan2011, Moulder2012}, superpositions of attractive and repulsive optical dipole potentials \cite{Marti2012}, and a painted potential  \cite{Ryu2013, Henderson2009}.  Separately, JJs for BECs have been realized with interference of laser beams \cite{Cataliotti2001, Albiez2005}, a magnetic trap with a repulsive barrier from a laser beam \cite{Levy2007}, and a radio frequency dressed magnetic trap \cite{LeBlanc2011}.  Recently \cite{Ramanathan2011, Wright2013}, a laser beam was used to create a barrier that acted as a weak link for atoms in a toroidal trap, the geometry of the rf atom SQUID, and induced phase slips.  In this Letter, we use the ``painted potential'' technique \cite{Henderson2009} to create and manipulate a BEC in a toroidal trap containing a pair of JJs, which is the cold atom analog of the dc SQUID.  In contrast to the weak link recently demonstrated in Ref.\,\cite{Wright2013}, the thin junctions reported here exhibit significant quantum tunneling.  Their behavior agrees with the predictions of the Josephson equations for ideal JJs with sinusoidal current-phase relation $i = i_c \sin \phi$, where the JJ current $i$ has maximum value equal to the critical current $i_c$, and $\phi$ is the phase difference across the JJ. 

\begin{figure}
\includegraphics[width=3.5in]{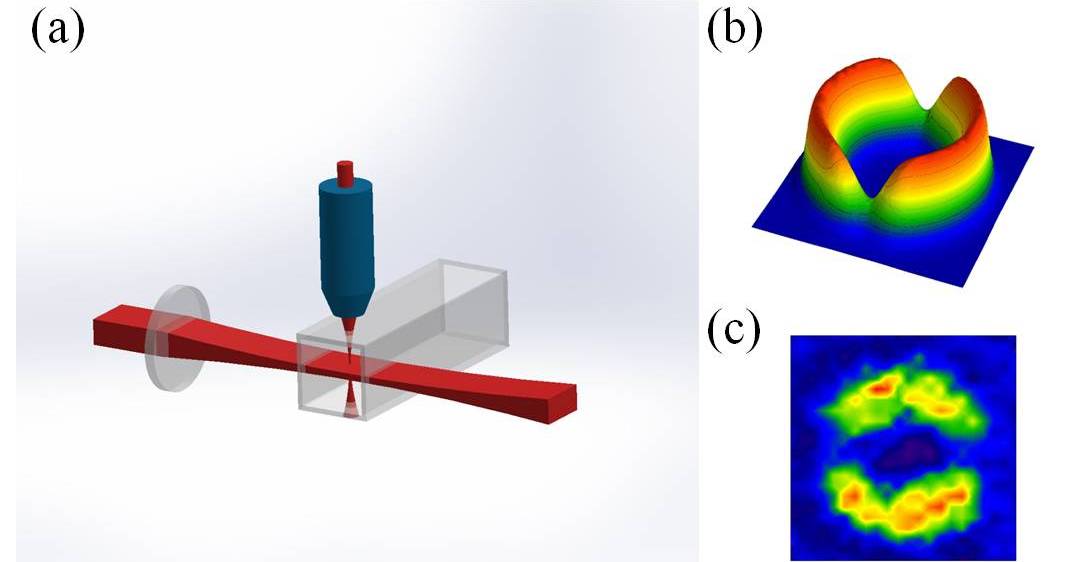}
\caption{\label{experiment}(a) Schematic of the experimental setup.  Arbitrary and dynamic potentials for the BEC are created by the superposition of two red-detuned beams making attractive optical dipole potentials.  One beam is a horizontal light sheet, the other is a rapidly moving focused beam which paints the potential. (b) Calculated intensity distribution of the 8\,$\mu$m diameter painted atom SQUID potential. (c) \textit{In situ} absorption image of a BEC in the atom SQUID potential. }
\end{figure}

The painted potential technique \cite{Ryu2013, Henderson2009} realizes complex dynamic potentials for BECs by rapidly moving far-detuned laser beams to create time-averaged optical dipole potentials.  Improvements to the setup described in Ref.\,\cite{Henderson2009} and used to demonstrate quantized circulation in a toroidal BEC \cite{Ryu2013}, particularly a new long working distance microscope objective (numerical aperture = 0.4), make the spatial resolution of the painted potential $\sim 1.5\,\mu$m.  That is comparable to the BEC healing length, so painted barriers can exhibit significant tunneling rates.  Figure\,\ref{experiment}(a) is a schematic drawing of the setup.  A horizontal scanning beam creates a flat two dimensional potential for supporting atoms against gravity, while the vertical painting beam creates a complex dynamic potential.  This high resolution system enables the painting of many varieties of complex and dynamic JJs, including the dc atom SQUID configuration of double JJs on a toroidal trap discussed here.  Figure\,\ref{experiment}(b) shows the calculated beam intensity distribution for this case and Fig.\,\ref{experiment}(c) shows the corresponding BEC.  The FWHM (full-width at half-maximum) of the barrier is 2\,$\mu$m, which is small enough for a BEC to have a significant tunneling rate (of order a few hundred hertz) for the typical density-weighted BEC healing length $\xi \sim 0.5\,\mu$m. 

\begin{figure}
\includegraphics[width=3.5in]{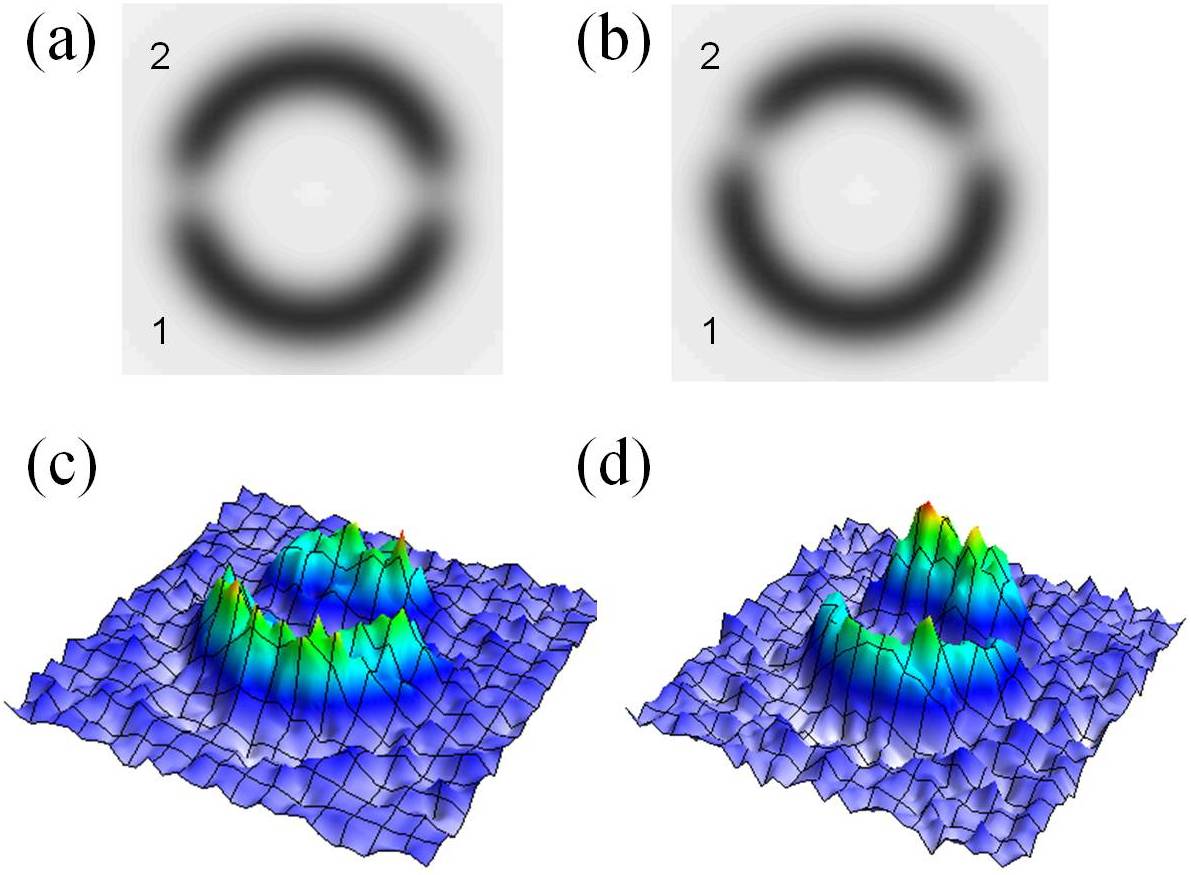}
\caption{\label{BECimages} Initial (a) and final (b) configuration of atom SQUID potential.  Images of a BEC where bias current is less (c) or more (d) than critical current.  Image (d) shows the BEC compression that occurs when tunneling is not possible.}
\end{figure}

It is hard to observe Josephson oscillations in this system directly because of their small amplitude.  Therefore, we used the scheme suggested in Ref.\,\cite{Giovanazzi2000} to study Josephson effects.  In our implementation of this idea, moving the JJs circumferentially towards each other leaves the BEC density unchanged as long as atoms can tunnel through the junctions to maintain the same chemical potential in both sectors of the torus.  This phenomenon of a current of atoms flowing without a chemical potential difference is the dc Josephson effect in this system.  The current will increase with barrier velocity until the critical current of the JJs is reached, at which point the system switches to the ac Josephson regime.  Here there is an oscillating current of atoms through the barrier, with frequency proportional to the chemical potential difference across the JJ, but no net current across it.  Therefore at barrier velocities greater than the speed limit imposed by the critical current, the moving JJs simply push the atoms, resulting in compression of atoms in one side and expansion in the other.  Figures\,\ref{BECimages}(a) and (b) show the initial and final potentials of the atom SQUID for the movement of JJs.  Figure\,\ref{BECimages}(c) shows the unchanged density distribution of atoms when atoms tunnel and Fig.\,\ref{BECimages}(d) shows the compressed cloud of atoms in one side when the JJs move too fast for tunneling. 

The simplest theoretical model for the experiment is based on the Josephson equations developed for a BEC \cite{Giovanazzi2000, Raghavan1999}.   Our symmetric system can be modeled as two condensates connected by a single JJ.  Letting $N_{1}, N_{2}$ and $\phi_{1}, \phi_{2}$ be the number and phase of atoms in the two sides of the torus, and defining the relative population difference $z = \left(N_{1}-N_{2}\right)/ \left(N_{1}+N_{2}\right)$ and the phase difference $\phi =\phi_{1}-\phi_{2}$, the Josephson equations are 
\begin{equation}
\dot z = I_{c} \sqrt{1-z^{2}} \sin{\phi}
\label{e:zdot}
\end{equation}
and
\begin{equation}
\dot \phi = -\omega_{C} \left(z-z_{0}\right)  - I_{c} \frac{z}{\sqrt{1-z^{2} }} \cos{\phi}.
\label{e:phidot}
\end{equation}
Here $I_{c}=2E_{J}/\hbar N$ is the critical normalized current of the double JJ, where $E_{J}$ is the Josephson coupling energy \cite{Giovanazzi2000, Ananikian2006} and $N=N_{1}+N_{2}$ is the total number of atoms.  Also, $\omega_{C}=E_{C}/2\hbar N$, where $E_{C}$ is the capacitance energy \cite{Giovanazzi2000, Ananikian2006}, and $z_{0}$ is the population difference when the chemical potential difference
\begin{equation}
\mu=\mu_{1}-\mu_{2}=\hbar \omega_{C} \left( z - z_{0} \right)
\label{e:mu}
\end{equation}
is zero ($\mu_{1}$ and $\mu_{2}$ are the chemical potentials in the two sides of the torus).  The derivative $\dot{z}$  is a normalized atom current, with maximum value $I_{c}$.  An effective bias current $\dot{z}_{0}$ is created by relative movement of the JJs against the BEC to change the equilibrium population difference $z_{0}$ \cite{Giovanazzi2000}.   The dc (ac) Josephson regime is $\dot{z}_{0} < I_{c}$ ($\dot{z}_{0} > I_{c}$).

We now discuss details of the experiment.  After evaporative cooling to 30\,$\mu$K in a quadrupole magnetic trap,  $^{87}$Rb atoms in the $F = 1$ and $m_{F} = -1$ state were transferred to the combined optical trap forming a toroidal potential with trapping frequencies of 300 and 570\,Hz in the vertical and radial directions, respectively.  The horizontal trapping beam, with wavelength $\lambda=1064$\,nm and waist $\omega_{0}=11$\,$\mu$m, was scanned horizontally at 14\,kHz over a width of 47\,$\mu$m.  The vertical trapping beam, with $\lambda=830$\,nm and $\omega_{0}=1.5\,\mu$m, painted (at a scan frequency of 25\,kHz) an 8\,$\mu$m diameter ring with two intensity minima forming JJs.  While the depth of the vertical painting potential remained fixed at 75\,nK, the horizontal trap depth was lowered from 40 to 1.1\,$\mu$K in 2.5 s to drive evaporation.  The result was a pure BEC with atom number $N$ set between 1000 and 8000, corresponding to chemical potentials from 29 to 58\,nK.  Interference patterns observed after a period of ballistic expansion confirmed that phase fluctuations in the thin BEC \cite{Petrov2001, Dettmer2001} were negligible.  Weak barriers (22\,nK) in the initial potential prevented spontaneous rotation of the atoms in the toroidal trap during evaporation.  At the end of the evaporation stage the barrier height was increased to 44\,nK in 100\,ms.   The two JJs were then moved towards each other at a rotation frequency $f$ for each junction, which gives rise to a bias current  $\dot{z}_{0} =4f$.  To prevent plasma oscillations and to establish the bias current adiabatically, the barriers were accelerated at constant rate to the desired constant velocity.  The total compression angle for each JJ was $\pi/8$ and the compression angle after the acceleration stage was $0.06 \pi$.  Following the movement of the JJs, the BEC was imaged by absorption \textit{in situ}.  The clear change seen in the final density distribution between the dc Josephson [Fig.\,\ref{BECimages}(c)] and ac Josephson [Fig.\,\ref{BECimages}(d)] regimes as the bias current increases can identify the critical current accurately.

\begin{figure}
\includegraphics[width=3.5in]{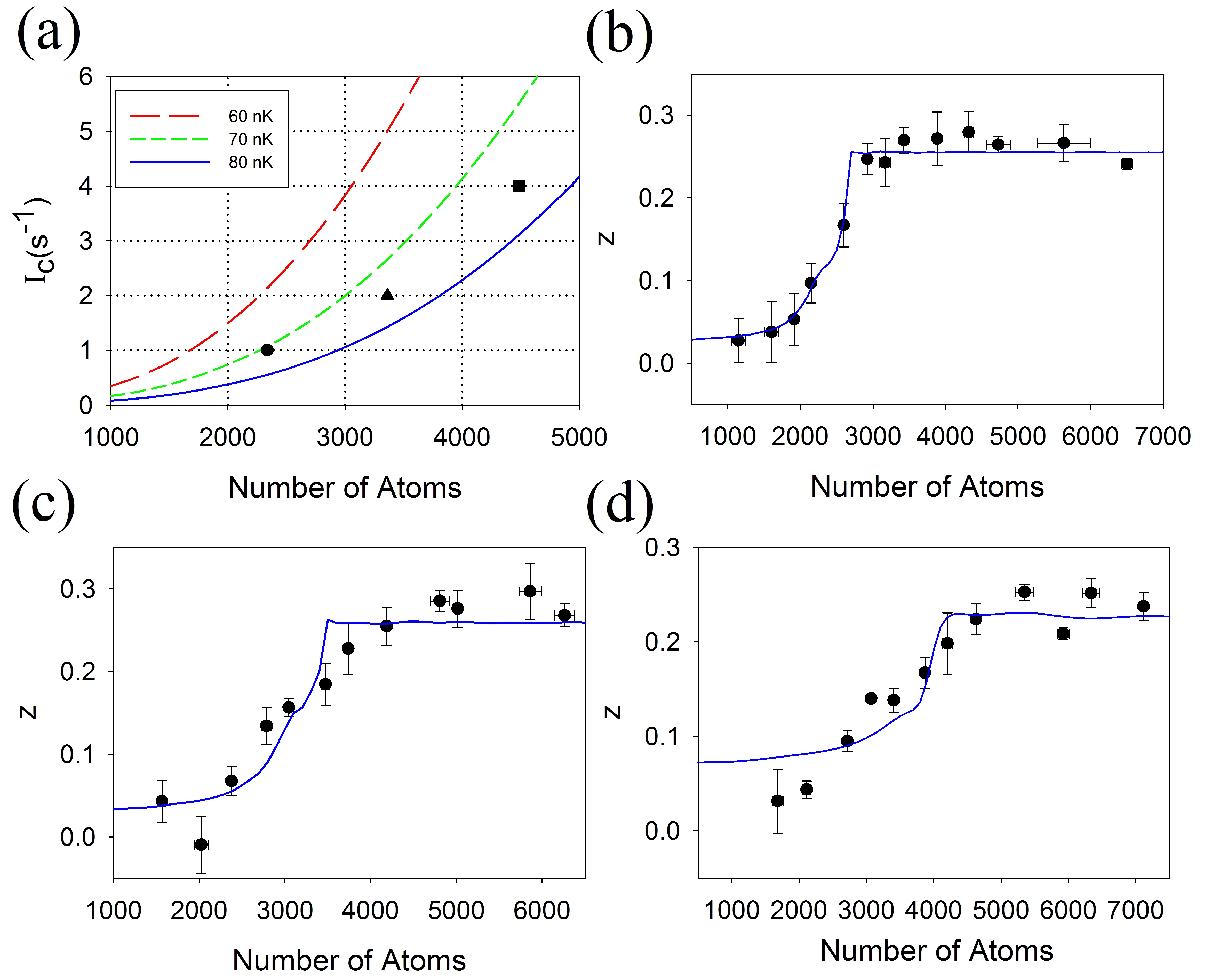}
\caption{\label{data} (a) Lines: computed critical current vs atom number for three trap depths.  Points show the measured critical atom numbers determined by the experiment (see text) with respective critical currents and fitted trap depth: 1\,s$^{-1}$ and 71\,nK (circle), 2\,s$^{-1}$ and 74.5\,nK (triangle), 4\,s$^{-1}$ and 75.6\,nK (square).  (b) - (d): Measurement of critical currents.  Each data point is the mean of several measurements of the final normalized population difference $z$, with the error bar being the standard error of the mean.  Systematic calibration uncertainty in atom number is less than $10\%$. Solid (blue) curves show the best fit solution of the Josephson equations.  The JJ rotation frequency $f$, corresponding bias current $\dot z$, and duration of JJ movement $T$ are: (b) $f=0.25$\,Hz, $\dot z=1$\,s$^{-1}$, $T=370$\,ms. (c) $f= 0.5$\,Hz, $\dot z=2$\,s$^{-1}$, $T=185$\,ms. (d) $f=1$\,Hz, $\dot z=4$\,s$^{-1}$, $T=92.5$\,ms.}
\end{figure}

A two-mode model that includes the nonlinear interaction in the coupling term \cite{Giovanazzi2000, Ananikian2006} predicted the critical currents plotted in Fig.\,\ref{data}(a) for three potential depths, showing that the critical current may be varied substantially over the range of atom numbers accessible to the experiment.  The critical current $I_{c}=2E_{J}/\hbar N$ depends on potential depth because $E_{J}$ is a function of barrier height, which is, here, a fixed fraction of the trap depth.

Data was collected by fixing the bias current and varying the atom number.  Values for $N$ and $z$ were obtained from the absorption images.  Figures\,\ref{data}(b)-(d) show three data sets corresponding to bias currents of 1\,s$^{-1}$, 2\,s$^{-1}$, and 4\,s$^{-1}$.  The sharp transition from dc to ac Josephson regimes is seen as the sudden change of $z$ as $N$ decreases.  The transition point shifts to higher $N$ as the bias current increases, in accordance with Fig.\,\ref{data}(a).  To make a quantitative comparison with theory we find the best least-squares fit to each data set of a numerical integration of the Josephson equations (1) and (2).  Each fit had three free parameters:  the potential depth, and the initial and final values of $z$.  These parameters are determined by the fit because slight deviations of the initial and final $z$ from the values predicted by geometry result from imperfections in the toroidal potential, and drifts in the relative waist positions of the horizontal and vertical trapping beams can introduce small variations in trap depth between data sets.  The fits, shown as the blue curves in Figs.\,\ref{data}(b)-(d), are in generally good agreement with the data.  The potential depths obtained from the fits (71, 74.5, and 75.6\,nK respectively) are within the range of 75(6)\,nK corresponding to the $\pm 2\,\mu$m estimate of the possible relative drift of the waists of the trapping beams and the uncertainty in the trapping beam intensity.  The critical atom numbers determined by the fit for each rotation frequency are shown on Fig.\,\ref{data}(a).  Their consistency with the theory predictions for the trap depths determined by the fits illustrates the agreement between experiment and theory already seen in Figs.\,\ref{data}(b)-(d).

\begin{figure}
\includegraphics[width=3.5in]{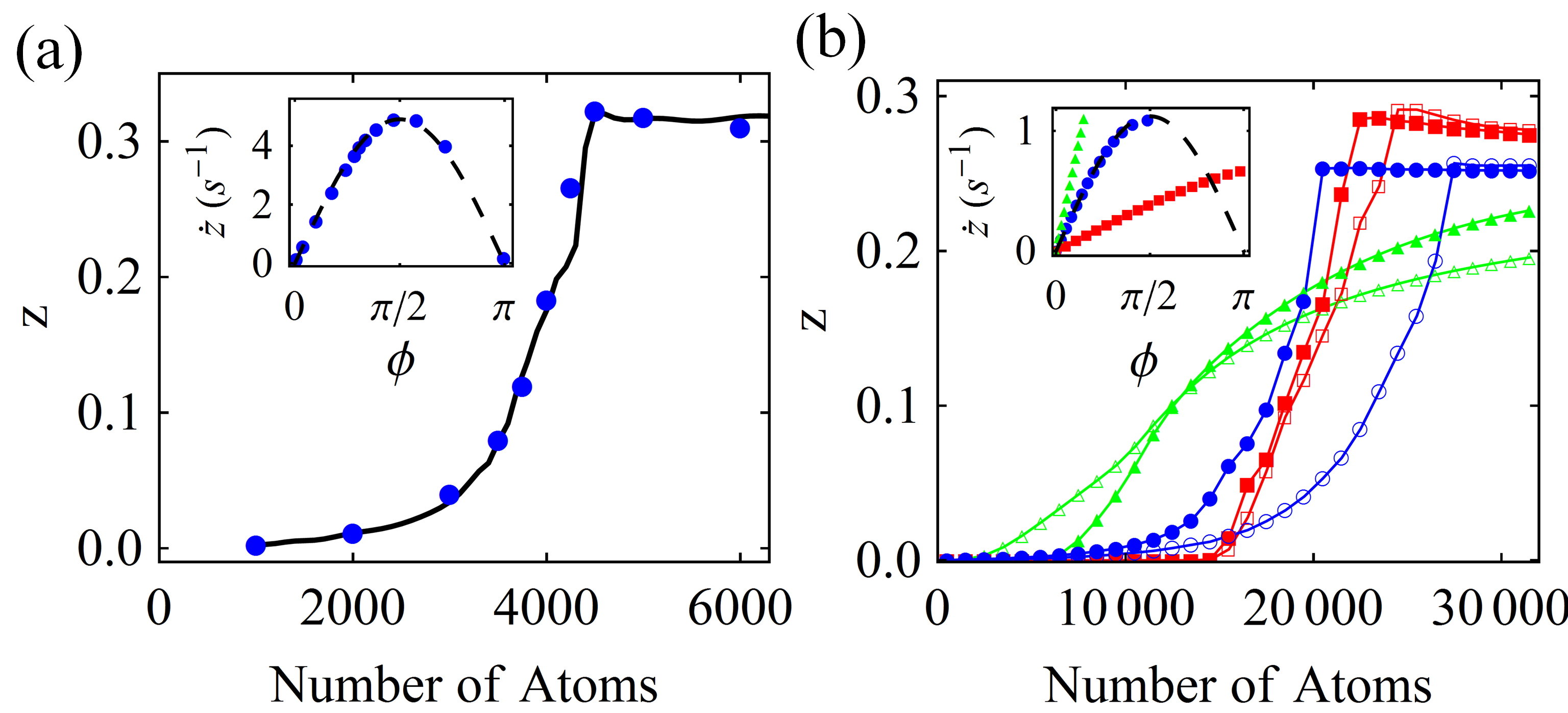}
\caption{\label{GPE}  (a)  Final value of normalized population difference $z$ for $f=1$\,Hz computed using the Josephson equations (solid line) and the 3D GPE (disks).  Inset:  disks: current-phase relation from GPE simulation for $N=4250$ atoms.  Dashed line:  ideal current-phase relation $\dot z =2E_{J}/\hbar N \sin{\phi}$. (b) Analogous 1D GPE simulations for a torus with radius $20\,\mu$m.  Barrier FWHM (in units of healing length $\xi$) and height (in units of chemical potential $\mu$) at $N=22\,000$ atoms: $3.9\xi$ and $1.8\mu$ (disks), $37\xi$ and $0.93\mu$ (squares), and $36\xi$ and $0.51\mu$ (triangles). Barrier rotation frequencies:  0.2\,Hz (solid disks and squares), 0.5\,Hz (open disks and squares), 5\,Hz (solid triangles), 12.5\,Hz (open triangles). Inset: Corresponding current-phase relations for $N = 22\,000$ atoms.  Dashed line:  ideal current-phase relation for the tunnel JJ.}
\end{figure}
The model based on the Josephson equations (\ref{e:zdot}) and (\ref{e:phidot}) is simple, assuming a sinusoidal current-phase relation and neglecting spatial effects such as condensate excitations and changes in $E_J$ and $E_C$ as the barriers move.  To understand the importance of these effects we performed three dimensional (3D) dynamic simulations of the experiment using the Gross-Pitaevskii equation (GPE).  Figure\,\ref{GPE}(a) shows that predictions of the Josephson equations are in excellent agreement with GPE results.  Further, the current-phase relation for the JJ obtained from the GPE solutions is sinusoidal and in excellent agreement with the prediction $\dot z =2E_{J}/\hbar N \sin{\phi}$.  This finding is consistent with theoretical studies of the current-phase relation for BEC flow through a 1D square barrier \cite{Watanabe2009, Piazza2010}.  The validity of the Josephson equations implies that the system reported here is analogous to the ideal dc SQUID, and so superfluid analogs of the phenomena seen in that device should be observable in the dc atom SQUID.  It also shows that the atom SQUID can be well understood from microscopic theories without relying on phenomenological models, simplifying design and interpretation in future atom SQUID research.

Since Josephson effects can be observed with both tunnel junctions and weak links, it is interesting to compare them in the contexts of the experiment reported here and of rotation sensing with atom SQUIDs.  Although experiments with superfluid helium have shown that sufficiently small weak links can have sinusoidal current-phase relations \cite{Sato2012}, analogous devices have not been realized for BECs, and so, we consider here weak links based on thick barriers (e.g \cite{Wright2013}).  Figure\,\ref{GPE}(b) shows GPE simulation results for a ring five times bigger than the experiment (to make room for a thick barrier) containing five times as many atoms (to leave the healing length and chemical potential unchanged).  The simulations show that a thick barrier weak link (squares) and our thin barrier tunnel junction (disks) would behave very differently in the experiment.  Specifically, as the atom number $N$ increases, we find that flow for the thick barrier starts when the chemical potential reaches the barrier height.  The system then passes through a narrow (in $N$) sinusoidal current-phase regime with very small critical current \cite{Piazza2010}.  For any higher atom number the thick barrier weak link has a nonsinusoidal current-phase relation and a critical velocity that increases with $N$ rapidly compared to the tunnel junction behavior shown in Fig.\,\ref{data}(a) \cite{Piazza2010}.  As a result, the final $z$ versus $N$ curves in Fig.\,\ref{GPE}(b) for the tunnel junction and the thick barrier weak link are quite different [the curve for a low thick barrier (triangles) is even more different because the critical velocity here is approaching the speed of sound].  More significantly, Fig.\,\ref{GPE}(b) shows that the critical atom number for thick barrier weak links is fairly insensitive to $f$ (because the critical velocity is very sensitive to $N$), while the tunneling regime of the thin barrier is distinguished by relatively large changes in critical atom number with $f$, as is seen in the experimental measurements (Fig.\,\ref{data}).  

An interesting next step would be to measure the junction's current-phase relation directly by, for example, allowing the two parts of the toroid to expand and overlap, forming interference fringes that indicate the relative phase across the junction.  Looking to the future, quantization of circulation will lead to quantum interference of currents in the dc atom SQUID as a function of rotation rate.  The interference can be seen via the shift of critical current with trap rotation rate, enabling the use of atom SQUIDs for rotation sensing.  In this case the relative insensitivity of the tunnel junction critical current to changes in $N$ should make it preferable to a thick barrier weak link.  The atom SQUID interaction region can be compact, so it will be interesting to compare its rotation-sensing performance with the current state of the art, the atom optics-based gyroscope \cite{Gustavson1997}.   The painted potential technique also allows for the creation of more complex circuit geometries, such as networks of rings \cite{Search2009}, and the system could also be used to study instability mechanisms leading to vortex formation when a BEC flows through a constriction  \cite{Piazza2011}.  It may be possible to apply the technology demonstrated here to recent proposals \cite{Nunnenkamp2008,  Hallwood2011, Schenke2011, Solenov2010, Solenov2010b, Solenov2011} to create macroscopic quantum superpositions of different flow states in an atom SQUID.  These states are analogous to the macroscopic superpositions of different flux states demonstrated in SQUIDS \cite{van der Wal2000,Friedman2000}.  There has been considerable discussion regarding the true number of entangled particles in these systems \cite{Marquardt2008, Korsbakken2010}, and so, it would be intriguing to duplicate the experiment with a BEC, where the microscopic physics is well understood.

In summary, we have created a pair of Josephson junctions on a toroidal Bose-Einstein condensate and demonstrated Josephson effects in a toroidal BEC by showing the transition between the dc and ac Josephson regimes and by measuring the critical current of the junctions.  The experimental data for this dc atom SQUID geometry is in good agreement with the predictions of the ideal Josephson equations and of the Gross-Pitaevskii equation, and the GPE simulation of the experiment finds that the JJ current-phase relation has the ideal sinusoidal form.

We gratefully acknowledge inspiring conversations with Eddy Timmermans and Dima Mozyrsky.  This work was supported by the U.S. Department of Energy through the LANL-LDRD program.

\end{document}